\documentclass[12pt]{article}

\usepackage{subcaption}
\usepackage{tikz,pgfplots}
\usepackage{amsmath, amsfonts, amssymb}
\usepackage[margin=1in]{geometry}
\usepackage{authblk}
\usepackage{natbib}

\title{Deep Autoassociative Neural Networks for Noise Reduction in Seismic data}

\author[1, 2]{Debjani Bhowmick}
\author[3]{Deepak K. Gupta}
\author[2]{Saumen Maiti}
\author[4]{Uma Shankar}

\affil[1]{\small Centre for Mine Planning and Design Institute, Kanke Road, Ranchi 834008, Jharkhand, India}
\affil[2]{Dept. of Applied Geophysics, Indian Institute of Technology, ISM Dhanbad 826004, Jharkhand, India}
\affil[3]{Dept. of Precision and Microsystems Engineering, Delft University of Technology, Mekelweg 2, Delft 2628CD, The Netherlands}
\affil[4]{Dept. of Geophysics, Banaras Hindu University, Varanasi 221005, Uttar Pradesh, India}
\date{}

\begin{document}

\maketitle

\section{Introduction}
\emph{Machine learning} is currently a trending topic in various science and engineering disciplines, and the field of geophysics is no  exception. With the advent of powerful computers, it is now possible to train the machine to learn complex patterns in the data, which may not be easily realized using the traditional methods. Among the various machine learning methods, the artificial neural networks (ANNs) have received enormous attention, and these methods have been used on various geophysical problems (\emph{e.g.} \cite{Poulton1992}, \cite{McCormack1993}, \cite{Bhowmick2016}, among others).  

The goal of most ANN formulations is to learn a mapping from input to output, and then be able to use the learnt mapping on another dataset. Among these, an autoassociative neural network (autoNN) tries to learn the reconstruction of input itself using backpropagation (\citealt{Kramer1992}). In an autoNN, the input and output are the same, and an approximation to the identity mapping is obtained in a nonlinear setting. AutoNNs have primarily been used to extract sparse internal representations of any input and reduce its dimensionality. \cite{Kramer1992} used a combination of linear and nonlinear activations, and used autoNNs for gross noise reduction in process measurements. However, with a shallow neural network, only limited improvements were observed.

\emph{Autoencoder}, a variant of autoNNs in a deep network framework, is a powerful technique to learn the internal representation of any input (\citealt{Bengio2009}). \cite{Vincent2010} used autoencoders for denoising tasks, however, these were primarily intended towards making autoNNs  more robust to noise. In a geophysical context, \cite{Valentine2012} used autoNNs for dimensionality reduction and quality assessment of seismic waveform data. In this work, building up over the idea of \cite{Kramer1992}, we use autoNNs in a deep network setting to reduce noise in geophysical data. In this paper, the first results of this study are presented. First, the method is discussed in brief, followed by the demonstration of deep autoNNs on a basic mathematical example and a geophysical example. 

\section{Theory}

\begin{figure}
    \centering
	\begin{tikzpicture}
    \node[anchor=south west,inner sep=0] at (0,0){
	\includegraphics[scale=0.6]{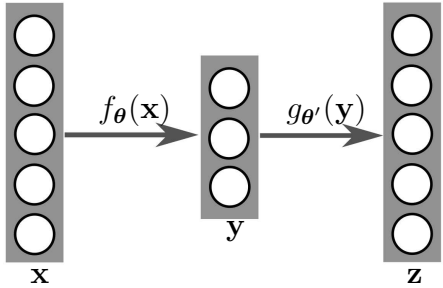}};
	\node[text width=2.5cm] at (2.6, 2.1) 
    {$f_{\boldsymbol\theta}(\mathbf{x})$};
    \node[text width=2.5cm] at (5.1, 2.1) 
    {$g_{\boldsymbol\theta'}(\mathbf{y})$};
	\node[text width=0.5cm] at (0.7, -0.1) 
    {$\mathbf{x}$};
    \node[text width=0.5cm] at (5.7, -0.1) 
    {$\mathbf{z}$};
    \node[text width=0.5cm] at (3.3, 0.5) 
    {$\mathbf{y}$};
	\end{tikzpicture}
    \caption{Schematic structure of a traditional autoassociate neural network.} 
    \label{fig_basic_ae}
\end{figure}

\indent A traditional autoNN consists of two parts: an \emph{encoder} and a \emph{decoder}. Fig. \ref{fig_basic_ae} shows the schematic diagram of a simple autoNN consisting of a single hidden layer. The mapping phase where the input $\mathbf{x}$ is transformed into the hidden representation $\mathbf{y}$ is termed as an encoder. A decoder is the part of an autoencoder where the input is reconstructed back $\mathbf{z}$ from its hidden representation $\mathbf{y}$. In Fig. \ref{fig_basic_ae}, the encoding and decoding functions are denoted by $f_{\boldsymbol\theta}(\cdot)$ and $g_{\boldsymbol\theta'}(\cdot)$, respectively and these mappings are parametrized by vectors $\theta$ and $\theta'$, respectively. Typically, the mapping functions comprise of affine mapping followed by certain nonlinearity and can be expressed as:
\begin{align}
& f_{\boldsymbol\theta}(\mathbf{x}) = \mathcal{S}(\mathbf{Wx + b}), \\
& g_{\boldsymbol\theta'}(\mathbf{y}) = \mathcal{S}(\mathbf{W'y + b'}),
\end{align}
where, $\mathbf{\theta} = \{\mathbf{W, b}\}$ and $\mathbf{\theta'} = \{\mathbf{W', b'}\}$ are parameter sets with $\mathbf{W}$ and $\mathbf{W'}$ denoting the weight matrices and $\mathbf{b}$ and $\mathbf{b'}$ representing the bias vectors, respectively. Typically the nonlinear mapping $\mathcal{S}(\cdot)$ is achieved using sigmoid or radial basis functions. 

The goal of the autoencoder presented in Fig. \ref{fig_basic_ae} is to minimize the reconstruction loss between $\mathbf{x}$ and $\mathbf{z}$. As error (loss) function, typically squared-error function or cross-entropy is used depending on the type of problem. The autoencoder presented in Fig. \ref{fig_basic_ae} consists of a single hidden layer. However, for autoencoders of higher complexity, several hidden layers can be used. Accordingly, the encoder and decoder will then comprise of a series of mappings in each. 

In the context of denoising, the input $\mathbf{x}$ in Fig. \ref{fig_basic_ae} is replaced with $\mathbf{\tilde{x}}$, where $\mathbf{\tilde{x}}$ comprises noisy as well as noisefree samples. Since the input contains noisy data and output does not, denoising autoNNs work in two folds: identify the primary signal and the random noise, and reproduce the primary signal. Although, non-coherent noise is used in this paper, the concept holds valid for coherent noise as well. For such cases, a deep autoNN needs to be used which can identify the the patterns in the primary signal and the coherent noise, and can separately reconstruct the noise as well (\emph{e.g.} for ambient-noise study).

\section{Examples}
\vspace{-0.5em}
\subsection{Mathematical example}
\vspace{-0.5em}
Here, we discuss the applicability of deep autoNNs for denoising tasks through an elementary example. A simple mathematical problem of 2 parameters ($z$ and $\theta$) is chosen and a process model of the following form is used (\citealt{Gupta2013}),
\begin{align}
& x = tz\sin\theta + \frac{t^2}{z}\cos\theta, \qquad t = 0.05:0.05:1, \nonumber \\
& z \in [0.5, 4.0], \qquad \theta \in [0.3, 1.3] \textrm{ in radians},
\end{align}
where, $t$ and $x$ are the input and output fields, respectively and $z$ and $\theta$ are the model parameters. For $t$, a range of 0.05 to 1 is chosen with a sampling interval of 0.05. The input field $\mathbf{t} = \{0.05, 0.1, \hdots, 1.0\}$ is then used to generate the output signal $\mathbf{x} = \{x_1, x_2, \hdots, x_n\}$, where $n$ denotes the number of sampling points. 

Once $\mathbf{x}$ is obtained, it is assumed that the process model is not known anymore. Next, an autoNN is trained to learn the internal representation of $\mathbf{x}$ such that for any noisy variant $\tilde{\mathbf{x}}$, the noise-free signal can be recovered. An autoNN comprising of 2 hidden layers is formulated. The number of neurons in each of the hidden layers is kept equal to $n$. Here, choosing the number of neurons equal to the number of input units does not lead to identity function due to the large noise added in some of the signals. Nonlinear (sigmodial) activation functions are used for projection from the input layer to hidden layer 1 and hidden layer 1 to hidden layer 2. For obtaining the output $\mathbf{z}$, linear  activations are used. 

A set of 20000 samples is used and is further divided into 80\% and 20\% for training and validation samples, respectively. Random noise of upto 25\% is added to the data point of 10000 samples and the other 10000 samples are kept noise-free. For 50\% of the noisy-samples, the noise scales based on the local value at the respective point of the sample. For the rest 50\%, the noise scales based on the mean of all the points of the sample. The gradient descent algorithm was used for optimization. 
\begin{figure}
\centering
	\begin{subfigure}{0.47\linewidth}
	\begin{center}
    \begin{tikzpicture}[scale = 0.75]
        \begin{axis}[%
        		ymin = 0,
        		thick, 
            axis x line=bottom,
            axis y line=left,
            legend pos = outer north east,
            xlabel = {test sample index},
            ylabel = {\% relative improvement ($\eta$)}
            ]
            \addplot[scatter, only marks, scatter src=\thisrow{class}, solid, thick, black] table[x=samples,y=improvement] {noise_reduc_dist_prob1_case1.dat};
        \end{axis}
    \end{tikzpicture}
    \caption{Relative noise reduction $\eta$ for the test samples}
	\end{center}
	\end{subfigure}
	\begin{subfigure}{0.47\linewidth}
	\begin{center}
    \begin{tikzpicture}[scale = 0.75]
        \begin{axis}[%
        		ymin = 0,
        		thick, 
            axis x line=bottom,
            axis y line=left,
            legend pos = outer north east,
            xlabel = {data point index},
            ylabel = {data point value $x$},
            legend pos = north west
			 ]
            \addplot[mark=square*, solid, thick, black] table[x=x,y=V] {prob1_case1_testsample_20.dat};
            \addplot[mark = triangle*, solid, thick, blue] table[x=x,y=Vnoisy] {prob1_case1_testsample_20.dat};
            \addplot[mark=*, solid, red, thick] table[x=x,y=Vae] {prob1_case1_testsample_20.dat};
                        
            \addlegendentry{noise-free}
			\addlegendentry{noisy}
			\addlegendentry{AutoNN corrected}
        \end{axis}
    \end{tikzpicture}
    \caption{Test sample 20}
	\end{center}
	\end{subfigure}
	\caption{(a) Relative noise reduction $\eta$ for the 100 noisy test samples, (b) noise-free, noisy and autoassociator corrected data for sample index 20. A denoising autoNN  comprising 2 hidden layers with 20 neurons in each is used.}
\label{fig_simp_ex}
\vspace{-0.5em}
\end{figure}
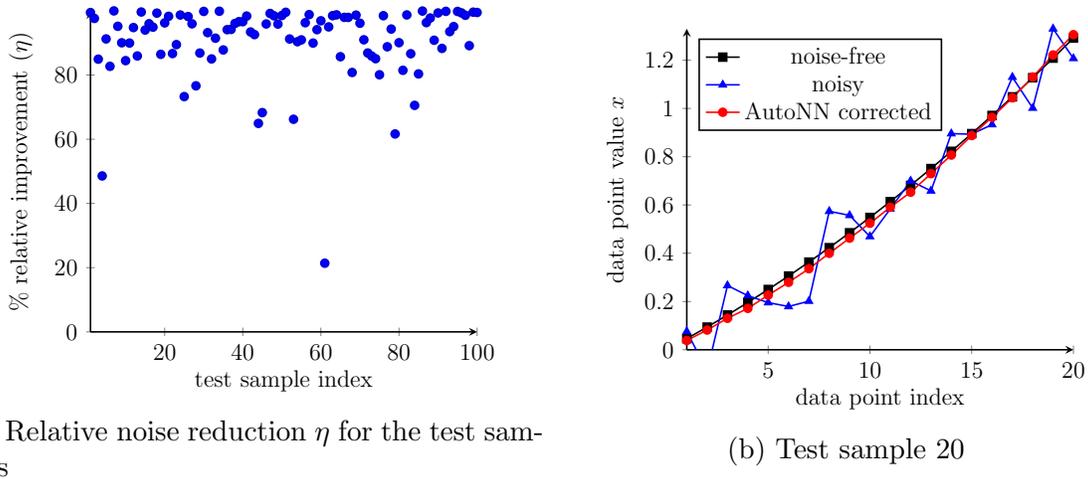

\indent To test the accuracy of the learnt representation $\boldsymbol\Phi$, 100 test samples are generated and a noise between 10\% and 25\% is added to each of these samples. Next, each of these samples $\mathbf{x}$ is passed through the learnt representation to reduce the noise and obtain the output $\mathbf{z} = \boldsymbol\Phi(\mathbf{\tilde{x}})$. The efficiency $\eta$ of the learnt representation $\boldsymbol\Phi$ for any noisy signal $\mathbf{\tilde{x}}$ is given by 
\begin{equation}
\eta  = 100 \times \frac{(\mathbf{z - x})^{\intercal}(\mathbf{z - x})}{(\mathbf{\tilde{x} - x})^{\intercal}(\mathbf{\tilde{x} - x})} \enskip \%.
\end{equation}
For the 100 test samples used, a mean value of $\eta = 90.27\%$ is obtained, which means that the learnt autoNN $\boldsymbol\Phi$ can reduce the noise by approximately 90\%. Fig. \ref{fig_simp_ex} shows the recovery value for each test example. This is a significant improvement and clearly demonstrates the applicability of autoNNs for denoising purpose. Fig. \ref{fig_simp_ex} also shows a randomly chosen sample (trace 20), where it can be seen that the trained autoNN can significantly remove noise from the data, and make it more interpretable.

\subsection{Denoising seismic data}
\begin{figure}[!htb]
   \centering
   \includegraphics[width=0.9\textwidth, trim = 0.25cm 0 0.35cm 0, clip=true]{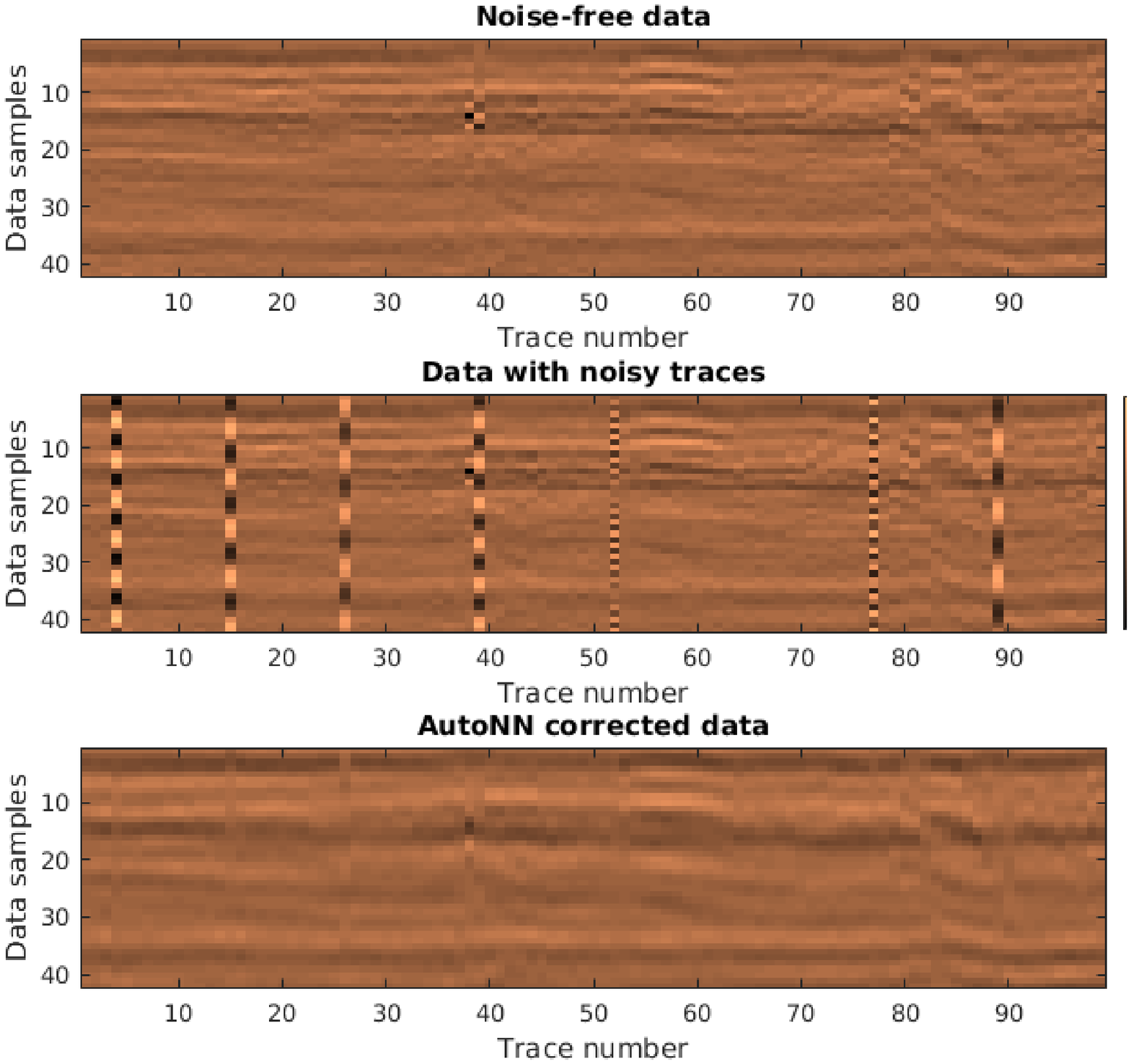}
   \caption{Noise-free seismic section (top), data with noisy traces added to it (middle), and output of autoassociative neural network model (bottom) trained for the removal of noisy traces local  	in time and frequency. The sampling interval is 8 ms, and the seismic section comprises 99 traces with 42 samples in each.}
   \label{fig_seismic1}
\end{figure}

Here, the application of denoising autoNN is demonstrated on seismic data corrupted with noise local in time and frequency. For testing purpose, a 2D seismic section comprising 99 traces and 42 time samples (at 8 ms interval) is used. To add synthetic noise, a few traces of the data are replaced with monofrequency sinusoidal traces. For the noisy traces, the amplitude is randomly chosen between \mbox{0.5$A_{\text{max}}$} and $A_{\text{max}}$, where $A_{\text{max}}$ denotes the maximum amplitude observed in the original noisefree seismic section. The frequency of the noisy traces is randomly chosen between 110 and 220 Hz. Next, an autoNN needs to be trained, such that the noise from test data can be removed.
 
To train the autoNN, approximately 0.38 million seismic images of size 9 traces $\times$ 42 samples are randomly chosen from the seismic section. For each sample, two noisy samples are generated by corrupting one of the traces. The amplitude and frequency ranges are kept to be the same as used in the test data. Thus, a total of 1.1 million training samples comprising noisefree as well as noisy images are used. Next a deep network comprising 3 hidden layers with 300, 400 and 300 neurons is used, and the autoNN is run for 50,000 epochs. The total dataset is split in a ratio of 0.95 and 0.05 for training and validation. 

Fig. \ref{fig_seismic1} shows the noisefree, noisy and autoNN correction results for the test data. From the seismic section shown in the middle, it can be seen that the data comprises of noisy traces. In the autoNN corrected data, significant reduction in noise is observed. However, it is seen that the designed autoNN regularizes the data due to which the resolution of data is reduced to a certain extent. An important aspect is that only high values of frequency (more than 100 Hz) are used to add noise to training as well as test data. The chosen training window is small, and cannot capture the behavior of low frequency noise. In future, we intend to adapt autoNN for other different types of noises including low frequency, multifrequency, \emph{etc}. Also, the a stacked version of autoNN needs to be tested on seismic data for lossless noise removal.    

\section{Conclusions}
In this paper, deep autoassociative neural networks (autoNNs) are discussed in the context of denoising of geophysical data. The applicability of autoNNs is demonstrated on a simple mathematical example, and first results of seismic data denoising are also presented. For the first example, different variants of autoNN are tested, and with stacked autoNN, more than 90\% reduction in noise is observed. For the test seismic data, it is observed that the autoNN can significantly remove the vertical time- and frequency-local noise. Future work includes testing larger examples with several different types of noise, and using deep-stacked-autoNNs to further reduce the noise, ensuring minimal compromise with the resolution of the signal.

\bibliography{cph18_start}

\end{document}